\documentclass[aps,prl,nofootinbib,superscriptaddress,longbibliography]{revtex4-1}

\usepackage{graphicx}
\DeclareGraphicsExtensions{.pdf}

\usepackage{amsfonts,amssymb,amsmath}
\usepackage{amsthm}
\usepackage{tikz}
\usetikzlibrary{calc}
\usetikzlibrary{arrows}
\usepackage{tikz-3dplot}
\usepackage{color} 
\usepackage{graphicx}

\newcommand{\comment}[1]{}

\theoremstyle{plain}

\theoremstyle{definition}

\begin{document}

\title{A Quantum Paradox of Choice and Purported Classical Analogues}

\author{Emily \surname{Adlam}} \affiliation{Centre for Quantum
  Information and Foundations, DAMTP, Centre for Mathematical
  Sciences, University of Cambridge, Wilberforce Road, Cambridge, CB3
  0WA, U.K.}  \author{Adrian \surname{Kent}} \affiliation{Centre for
  Quantum Information and Foundations, DAMTP, Centre for Mathematical
  Sciences, University of Cambridge, Wilberforce Road, Cambridge, CB3
  0WA, U.K.}  \affiliation{Perimeter Institute for Theoretical
  Physics, 31 Caroline Street North, Waterloo, ON N2L 2Y5, Canada.}

\date{September 2015}

\begin{abstract}
We recently considered the task of summoning an unknown quantum 
state and proved necessary and sufficient
conditions for Alice to be able to guarantee to complete
the task when there may be several possible calls, of which
she need only respond to one.   We showed that these are 
strictly stronger conditions than those previously established
by Hayden and May for the case where Alice knows
there will only be one call.  
We introduced the concept of a {\it quantum
paradox of choice} to summarize the implications of these results:
Alice is given more options to complete our version of the task, yet
one can easily construct examples where our version is impossible and
the apparently simpler version considered by Hayden-May is possible. 

Finkelstein has argued that one can identify analogous classical 
paradoxes of choice in a relativistic setting.   We examine Finkelstein's
proposed classical tasks and explain why they seem to us disanalogous.      
\end{abstract}

\maketitle

\section{Introduction} 

We introduced a recent paper \cite{akqpoc} with the following metaphor: 

``A Holistic Magician (HM) repeatedly performs the following trick.
He first asks you to give him an object that you are 
sure he cannot copy. 
After working behind a curtain, he presents you with 
$N$ boxes and asks you to choose one. 
Opening your chosen box, he reveals the original object inside. 

You initially imagine that he has arranged some concealed mechanism
that somehow passes the object sequentially through the boxes,
allowing him to stop the mechanism and keep the object in one box if
you select it.  However, you are then puzzled to notice that he is
unable to make the trick work if you select more than one box, even
though you allow him to choose which of your selections to open.  This
argues against your mechanical explanation, and indeed seems to make
any simple explanation problematic.  How can giving the magician more
freedom make him unable to complete the task?'' 

A comment by Finkelstein \cite{fcpoc} inspires another version:

``A Faux-magician (F) repeatedly performs the following trick.
He first describes a signal that can easily be 
generated and copied: a light flash, for example.    
After working behind a curtain, he presents you with 
$N$ boxes with buttons and lights and asks you to choose
one box and press its button.  When you do so, its light
flashes, and none of the other lights also flash. 
He stresses that he requires both of these conditions for the trick to 
have succeeded.    

You are not at all intrigued or puzzled when F tells you that the
trick (as he defines it) will not work if you press several buttons.
This is because there seems to be an obvious explanation: each button
is a switch for the corresponding light.  Pressing several buttons
would then cause several lights to flash, meaning that the trick as
defined would not work.  You thus tell F to get a better act and book
HM for your next party.''

Notice that the two magicians' tricks have something in common. 
In both cases you give them more ways of completing the trick by
pressing several buttons, and in both cases this appears to 
prevent them from completing the task.   But in the first case,
this seems somewhat surprising, while in the second, it seems very 
unsurprising.   The second story replicates one feature of the first while 
neglecting others.  As a result, we think, no one 
is likely to call the second story puzzling or paradoxical.   

\section{Recap} 

In Ref. \cite{akqpoc}, we considered the task of 
summoning \cite{Kentnosummoning}
an unknown quantum state.   This task involves two agencies,
Alice and Bob, who may have collaborating agents distributed
throughout space-time.   Bob secretly prepares a quantum
state, and he (i.e. his local agent) hands it over to Alice
(i.e. her local agent) at some point $s$ in space-time.   
Alice and Bob have agreed on a number of call points $c_i$
and corresponding return points $r_i$.   Bob may request the
state at any call point $c_i$, and Alice is then supposed
to return it at the corresponding $r_i$.    In the most 
interesting version of the task, $r_i > c_i$ and $r_i > s$ 
for each $i$, where $>$ denotes the causal relation between
space-time points.    We proved necessary and sufficient
conditions for Alice to be able to guarantee to complete
the task when there may be several possible calls, of which
she need only respond to one.   We showed that these are 
strictly stronger conditions than those previously established
by Hayden and May \cite{Hayden} for the case where Alice knows
there will only be one call.    We introduced the concept of a {\it quantum
paradox of choice} to summarize the implications of these results:
Alice is given more options to complete our version of the task, yet
one can easily construct examples where our version is impossible and
the apparently simpler version considered by Hayden-May is possible. 

As we noted in Ref. \cite{akqpoc}, the discussion in 
that paper follows the tradition of using parables and apparent
paradoxes to refine our understanding of quantum theory \cite{Aharonov,
cat, Cheshire, Zeno, Gibbsparadox, pigeon,lsw}.  The results of Refs. \cite{Hayden,akqpoc} rely
on relativistic causality as well as quantum theory, and we 
suggested in Ref. \cite{akqpoc} that the apparent tension between them 
may be the first intrinsically relativistic quantum paradox.    

There is also a counter-tradition (e.g. \cite{fthreebox,ferriecombes}
of criticizing these parables and paradoxes,
usually on the grounds that they do not seem (to the critics)
even superficially paradoxical, or that they are not   
intrinsically quantum theoretic, or both. 
This too can be interesting and valuable: it is 
certainly worth reflecting on exactly  
what any proposed example, such as ours, really teaches us 
about physical principles.    
As we understand it, Finkelstein \cite{fcpoc} follows this latter
tradition by suggesting that there are precise 
classical analogues of our quantum paradox of choice. 

\section{Finkelstein's example}

Finkelstein suggests the following purported classical analogue 
of our quantum paradox of choice (\cite{fcpoc}, emphasis added): 

``Here is a simple example, for which quantum restrictions are
not needed:
Say there are two laboratories called $L$ and $R$; let $D$ be the distance between
them, and $T = R/c$ the time required for a signal traveling at the speed of
light to go either from $L$ to $R$ or from $R$ to $L$ (all times and distances as
measured in the frame in which both $L$ and $R$ are at rest). There are two
tasks which $B$ might request of $A$:

{\bf Task 1} \qquad 
Send a signal from $L$ to arrive at $R$ at time $T$, but do
not send any signal from $R$ to $L$. If this task is requested, the
request is submitted to $A$ in laboratory $L$ at time $t=0$.

{\bf Task 2} \qquad 
Send a signal from $R$ to arrive at $L$ at time $T$, but do
not send any signal from $L$ to $R$. If this task is requested, the
request is submitted to $A$ in laboratory $R$ at time $t=0$.
Clearly it would not be possible for both
requests to be fulfilled (just as in
the [Adlam-Kent] example where the no-cloning restriction prevents more than one
request from being fulfilled). The situation appears paradoxical because B,
when making both requests, would be satisfied if
either one were fulfilled.'' 

Now, as stated, this example does not work, without imposing
further restrictions on $A$.   In particular, it does not work
in the framework we consider \cite{Kentnosummoning,qtasks} in which
she may coordinate a network of collaborating agents distributed 
wherever she chooses in space-time.   If $A$ is allowed this power,
she may station an 
agent $A_c$ on a line between laboratories $L$ and $R$, and send
all signals via this agent, who may instantaneously relay
them.   If $B$ makes both requests, then $A_c$ receives
two signals, and can choose to relay the first (or, if she is
at the midpoint, may choose to relay either one) and intercept
the other.   $B$ thus receives precisely one valid signal, at one
of the two laboratories, whether he makes one request or two. 

\section{Refining Finkelstein's example} 

However, Finkelstein's underlying point is clear, and the example can
be simply refined to make it work in our framework.   
Suppose that $A$ has two agents, $A_0$ and $A_1$, at spatially well separated
sites distance $D$ apart, and $B$ has agents $B_0$ adjacent to $A_0$ and $B_1$
adjacent to $A_1$, so that each $B_i$ is separated from $A_i$ by
distance
$\epsilon \ll D$.    All of these agents are stationary in some
mutually agreed intertial frame.   Now define the following non-local task. 
At time $t=0$ in the agreed frame, each $B_i$ will 
send the corresponding $A_i$ a
classical bit, $0$ or $1$.   $A$ is guaranteed that at least one
$1$ will be sent.    Her task is to return to the $B_i$, by time 
$t= 2 \epsilon$, two classical bits, one $0$ and one $1$, ensuring
that the $1$ is sent to an agent $B_i$ who sent her a $1$.  

Now, if $A$ were also guaranteed that only one $1$ will be sent, 
the task is trivial: each $A_i$ simply needs to return the bit they
are sent -- and this is the only way of satisfying the task.   
However, if it is possible that two $1$'s will be sent, then $A$
has no way of ensuring that she completes the task.
This is true although, if $A$ receives two $1$'s, there are two
possible valid ways of completing the task.     

\section{Discussion}

In the last example, $A$ may have more valid ways of completing the 
task, if two $1$'s are sent, but nonetheless this possibility 
makes it impossible for her to guarantee completion of the task.
But is this in any way {\it paradoxical}?    Successfully completing
the task involves returning to the $B_i$ appropriately anti-correlated
bits at space-like separated points.  If the $B_i$ promise to supply 
these bits -- in other words, 
if they promise to give the $A_i$ the data that complete
the task -- then the $A_i$ can indeed complete the task.   If they do
not, then the $A_i$ can not.   
We understand the term paradox to imply a challenge to
pre-existing intuitions, and we suspect few readers will feel
such a challenge here.   

Compare the summoning task in our original discussion \cite{akqpoc}. 
There Alice needs to get a single quantum state to 
some requested point in space-time.   
Yet it turns out that this is
strictly harder if she may be given several options for return
points.  That is, there are strictly fewer sets of request and return
points for which it can be achieved. 
This seems to us, and at least to 
some others with whom we have discussed the problem, an 
interesting and initially surprising feature of relativistic quantum
theory.   As discussed in Ref. \cite{akqpoc}, it challenges 
our understanding of whether and how quantum states can be localized
in space-time. 
 
Of course, the results are explicable.
The relevant theorems were proven
in Refs. \cite{Hayden} and \cite{akqpoc}, and 
the paradox was resolved as well as presented in Ref.
\cite{akqpoc}.  To summarize: it turns out that,  
in the version of the task where only one summons is allowed,
the anti-correlated data given to Alice by the summonses constitute
an exploitable resource.    However, to understand how and why 
this resource is relevant seems at present to require 
an intuitive understanding both of the possibilities
given by 
iterative uses of teleportation and quantum secret sharing
for transmitting 
quantum information in space-time  and of the constraints implied by
iterative uses of the no-signalling principle \cite{Hayden,akqpoc}.    

It is perhaps worth emphasizing that, in our view, not every situation
(classical or quantum) in which more options make a task harder 
deserves to be termed a paradox of choice.   A single path from
$A$ to $B$ is straightforward to navigate, however twisty it may be.    
Extending it into a maze generally makes things harder, even 
when there are several paths through the maze. 
But we would not call this a paradox: to deserve that term
requires a challenge to the intuition, and we see none
here. 
 
That said, it should also be acknowledged that scientific paradoxes
can only be characterized in terms of the limitations of human
cognition and of pre-existing mental models, and these may perhaps be
different for different readers.  
There may perhaps be
readers whose pre-existing intuitions about summoning relativistic
quantum information assured them that the precise results of
Refs. \cite{Hayden,akqpoc} must be true: if so, we salute them.
Although we think it a bit less likely, there may perhaps also be
readers whose pre-existing intuitions about classical information in
space-time told them that the refined Finkelstein example above should
not work.  For any such readers (but, we would say, only for them),
Finkelstein's term ``classical paradox of choice'' would indeed be
appropriate.

For us, the key point of Ref. \cite{akqpoc} is that considering
summoning quantum states with single and multiple calls reveals a new
and surprising physical distinction between the two, encapsulated in 
the quantum paradox of choice described therein.    This feature is  
intrinsic to relativistic quantum theory: nothing like it arises in
summoning quantum states in non-relativistic quantum mechanics, nor in 
summoning classical states in relativistic classical mechanics.

{\bf Acknowledgments} \qquad This work was 
partially supported by an FQXi grant and by
Perimeter Institute for Theoretical Physics. Research at Perimeter
Institute is supported by the Government of Canada through Industry
Canada and by the Province of Ontario through the Ministry of Research
and Innovation.   


%

\end{document}